\newcount\vertou
\ifnum\vertou=0
\documentclass[nofootinbib,floats,aps,superscriptaddress,preprint]{revtex4}
\usepackage{epsfig}
\def\myabs#1{\begin{abstract}\vspace*{4mm}#1\end{abstract}}
\def\mypreprint#1{\preprint{#1}\vspace*{1cm}}
\def\mark#1{}

\else
\documentstyle[times,pramana,epsf,psfig,floats]{ias}
\def\mypreprint#1{}
\newcommand\myabs\abstract
\fi
\begin{document}
\mark{{Realistic construction...}{Yuval Grossman and Gilad Perez}}
\mypreprint{\vbox{
    %\hbox{WIS/XX/03-MAR-DPP~~}
    \hbox{hep-ph/0303243}\hbox{March, 2003}}}
\title{Realistic split fermion models\footnote{%
\ifnum\vertou=0
Talk presented by Gilad Perez at PASCOS '03, 9th International
Symposium on Particles, Strings and Cosmology, January 2003, Mumbai,
India.
\else
Presented by Gilad Perez
\fi
}}
%%%%%%%%%
\ifnum\vertou=0
\author{Yuval Grossman}  
\affiliation{Department of Physics,  
Technion--Israel Institute of Technology,\\ 
Technion City, 32000 Haifa, Israel\vspace{6pt}}
\author{Gilad Perez}
\affiliation{Department of Particle Physics, Weizmann Institute of Science\\
 Rehovot 76100, Israel\vspace*{18pt}}
\else
\author{Yuval Grossman$^1$ and Gilad Perez$^{2}$}
\address{$^1\,$Department of Physics,  
Technion--Israel Institute of Technology, 32000 Haifa, Israel.\\ 
$^2\,$Department of Particle Physics, 
Weizmann Institute of Science, Rehovot 76100, Israel.}
\fi
%%%%%%%%%%
%%%%%%%%%%
%%%%%%%%%%
\myabs{The standard model flavor structure can be explained in
theories where the fermions are localized on different points in a
compact extra dimension. We explain how models with two bulk scalars
compactified on an orbifold can produce such separations in a natural
way.  We show that, generically, models of Gaussian overlaps are
unnatural since they require very large Yukawa couplings between the
fermions and the bulk scalars.  We present a two scalar model that
accounts naturally for the quark flavor parameters and in particular
yields order one CP violating phase.}

\maketitle

%%%%%%%%%%%%%%%%%%%%%%%%%%%%%%%%%
%%%%%%%%%%%%%   I   %%%%%%%%%%%%%
%%%%%%%%%%%%%%%%%%%%%%%%%%%%%%%%%

\begin{figure}[tbp]
\ifnum\vertou=0
\epsfxsize=11.3cm
\else
\epsfxsize=8.42cm
\fi
  \centerline{\epsfbox{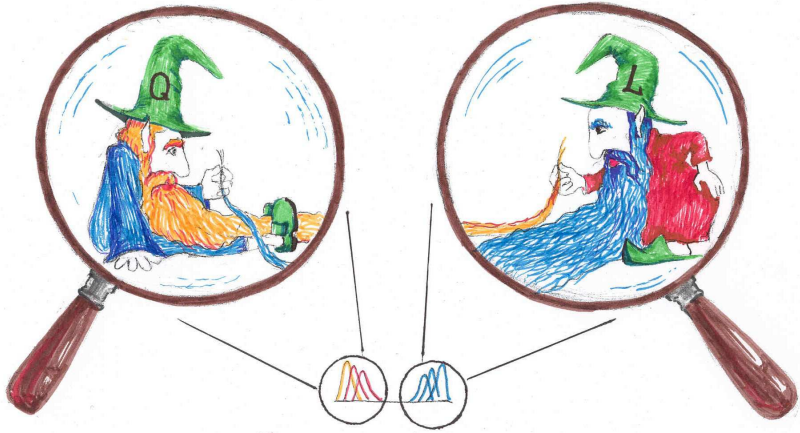}}
\caption{Configuration of the fifth dimension
 fermion wave functions in the AS model. Tiny overlaps between the
 quark (reddish curves on the LHS) and the leptons (blueish curves on
 the RHS) prevent proton decay.  Small overlaps between the quarks
 yields the flavor hierarchy.}
\label{overlap}
\end{figure}

The split fermion framework of Arkani-Hamed and Schmaltz (AS)
\cite{AS} provides a possible explanation for the flavor
puzzle.  It is based on the idea that the standard model  fermions
are separated in one or more extra dimensions. Consequently, the four
dimensional (4D) Yukawa couplings between fermions are naturally
suppressed by the overlap of their wave functions.  Proton stability
may be also accounted for in such framework provided that the quarks and
leptons are well separated in the extra dimensions, see
fig. \ref{overlap}.  

Despite the attractiveness of this framework there is no complete
realization of it.  The two main requirements from any realization are
chirality of the induced low energy 4D model, and separation of the
fermion wave functions in the extra dimensions.  In \cite{AS} a five
dimension (5D) model was presented.  It uses domain wall fermions,
namely, a bulk scalar field with non trivial vev that couples to the
fermions.  In addition, the fermions have different bulk masses in
order to get the required separation between the fermion wave
functions in the fifth dimension.  A specific configuration for this
model was presented in \cite{MS}. Assuming that the fermion wave
functions are Gaussian, it produced the observed fermion masses and
mixing angles.  The model is chiral, however, only in the limit of an
infinite extra dimension. This is not satisfactory since in reality
the extra dimension must be finite.  For a finite extra dimension a
chiral theory can be obtained if the fifth dimension is an orbifold
where the orbifold symmetry is used to project out one of the zero
modes~\cite{GGH,KT}. In such scenarios, however, the orbifold symmetry
also forbids bulk masses.  This is problematic since different bulk
masses are needed in order to split the different fermions. Indeed, in
that case the zero modes are localized at the one of the
orbifold fixed points, and the fermions rich flavor structure cannot
be naturally accounted for.

We consider a model with two scalar fields that couple to the fermions
\cite{GrP}.  This rather minor modification of the ideas presented in
\cite{GGH,KT} can produce fermion localization in the bulk.  The advantage
that two scalar models have over one scalar models is that the
effective mass can vanish in the bulk.  Intuitively the picture is as
follows. With one bulk scalar the sign of the Yukawa coupling between
the fermion and the scalar determines the boundary where the fermion
is localized \cite{GGH}. Once a second scalar with opposite sign
Yukawa coupling is introduced, the picture is more complicated.  The
second scalar tends to localize the fermion on the other boundary.
Sometimes, one scalar is dominant and the boundary where the fermion
is localized is determined by the sign of the coupling between
the fermion and the dominant scalar.  In other cases, however, the
tension between the two scalars results in a compromise: A
configuration where the fermion is localized in the bulk.

Below we explain how two scalar models can naturally account for the
quark masses, mixing angles and CP violating phase \cite{GrP}.  In
particular, in such  models the fermion wave functions are not of
Gaussian shape. This implies that the resultant 4D mass matrices do
not contain very small entries. The fact that there are no small
entries ensures large CP violating phase, as required. This is in
contrast to models where the fermion wave function are assumed to be
Gaussian
\cite{AS,MS,Branco} where the  CP violation phase is very small.
In fact, there are more problems for realistic models with Gaussian
wave functions.  We found that they are generically unnatural. For
example, in the above framework proton stability requires fine tuning
of ${\cal O}\left(10^{-4}\right)$.

We now move to present the model of \cite{GrP}.  The space-time of the
model is described by an $M_4\times S_1/Z_2$ orbifold. The physical
region is defined as $0 \le u \le 1$ where $u\equiv x_5/L$ such that
$L$ is the size of the extra dimension. The model includes the
standard model fermions $\psi^j$ and two scalars, $\varphi_i$
($i=1,2$) which acquire vevs $v_i$.  For each $j$ one of the fermion
components, $\psi^j_{R}$ or $\psi^j_L$, is even 
while the scalars and the other component of the fermions are odd
under the orbifold symmetry.  We assume that there is no interaction
between the two scalars. In a scaling where
$[\psi^j]=[v_i]=\,3/2$ and $[\varphi_i]=[a_i]=[f]=[X]=\,0$
the relevant part of the Lagrangian can be written as \cite{GrP}
\begin{eqnarray}{\cal L}_{\rm int}&=&
-{1\over L^2}\left[
fa_1\bar\psi \left(
\varphi_1-X\varphi_2\right)\psi+
\sum_i v^2_i {a^2_i\over2}\left(\varphi^2_i-1\right)^2\right]
\,, \label{L5D2S}
\end{eqnarray}
where we suppressed the flavor indices and
ordered the scalars such that  $a_2/a_1 > 1$. 
In the large $a_i$ limit  
the scalar vevs are approximated by \cite{GP} 
\begin{equation}\label{hi}
h_i(u)\equiv\langle\varphi_i\rangle(u)=
\tanh\left[a_i u\right] \tanh\left[a_i (1-u)\right] \,.
\end{equation}
The configuration of the fermions are determined by
the following function
\begin{equation}\label{gu}
g(u)= h_1(u) - X h_2(u)\,,
\end{equation}
which plays the role of an effective bulk mass, see eq. (\ref{L5D2S}).
The value of $X$ is very important.  When $X < a_1/a_2$ or $X > 1$ the
function $g(u)$ vanishes only at the orbifold fixed points. For
$a_1/a_2<X<1$, however, $g(u)$ has four roots where two of them are in
the bulk.  Since the fermions are localized around the roots of
$g(u)$, in that case they can be localized also in the bulk and not
only at the orbifold fixed points. This property is important since it
allows for separation between different fermions.

The wave function of the fermion zero mode (see e.g. \cite{AS})
is given by:
\begin{equation} \label{solkx}
y(u)\propto\exp\left[-fa_1\int_0^{u}\, g(w)\, dw \right]\,.
\label{psi5}
\end{equation} 
The local maxima of $y(u)$ are at the roots of $g(u)$ which we denote
by ${u_{\rm max}}$. In general there can be up to two maxima for
$g(u)$. The dominant maximum can be at one of the orbifold fixed
points or in the bulk. Moreover, there are cases where the two maxima
are significant. 

To anylize quark masses and mixing we need to restore
the flavor indices of the model. Then, $f$
and $X$ are promoted to be flavor matrices.  Based on the above
analysis we constructed an example of a configuration which
reproduces the correct quark masses mixing angles and CP violating
phase \cite{GrP}.

Beside addressing the flavor problem, split fermion models can also be
used to ensure long enough lifetime for the proton.  For that, the
quarks and the leptons 5D wave functions should be localized far away
from each other with roughly Gaussian profiles \cite{AS}. This is
translated to the requirement that the maximum of the wave function,
${u_{\rm max}}$, must be within the linear region of $h_i(u)$, see
eq.  (\ref{hi}), namely at
${u_{\rm max}}\ll 1/a_1$.  For concreteness we shall take
${u_{\rm max}}\sim 0.3/a_1$.  The width of the wave function in the
linear region is given by $\Gamma^{-1}\sim \sqrt{fa^2_1}$. In order
that this model will correctly reproduce the quark flavor parameters
the following relation should hold \cite{MS}
\begin{eqnarray}
\Gamma^{-1}{u_{\rm max}} \sim 0.3\,\sqrt{f\over2}\sim18\ \ \Longrightarrow \
f\sim10^4
\label{f1}\,.
\end{eqnarray}
This result is independent of the size of the extra dimension, $L$.
Thus, it is generic that in realistic models
proton stability requires fine tuning of 
${\cal O}\left(10^{-4}\right)$.
%%%%%%%%%%%%%%%%%%%%%%%%%%%%%%%%%
%%%%%%%%%%%%%  Ack  %%%%%%%%%%%%%
%%%%%%%%%%%%%%%%%%%%%%%%%%%%%%%%%
\ifnum\vertou=0
\acknowledgments 
We  thank the organizers of the PASCOS-03 conference at the
TIFR for their hospitality and for a great conference.  We thank Yossi
Nir and Yael Shadmi for useful discussion and comments on the
manuscript.  GP acknowledge the support from the Clore Scholars
Programme.
\fi

\end{document}